\documentclass[preprintnumbers, floatfix, 
preprintnumbers, letterpaper, superscriptaddress,nofootinbib, twocolumn]{revtex4}

\usepackage{graphicx}
\usepackage{microtype}
\usepackage{amsmath}
\usepackage{amssymb}
\usepackage{subfigure}
\usepackage{hyperref}
\usepackage{url}
\usepackage{xcolor}
\usepackage{color}
\usepackage{mathrsfs}
\usepackage{calrsfs}
\usepackage{amsfonts}
\usepackage{latexsym}
\usepackage{ragged2e}
\usepackage{epsfig}
\usepackage{textcomp}

\usepackage{caption}
\DeclareCaptionJustification{justified}{\leftskip=0pt \rightskip=0pt \parfillskip=0pt plus 1fil}
\definecolor{vividviolet}{rgb}{0.62, 0.0, 1.0}
\definecolor{amaranth}{rgb}{0.9, 0.17, 0.31}
\definecolor{palatinateblue}{rgb}{0.15, 0.23, 0.89}
\definecolor{brightpink}{rgb}{1.0, 0.0, 0.5}
\definecolor{cornflowerblue}{rgb}{0.39, 0.58, 0.93}
\definecolor{deepcarminepink}{rgb}{0.94, 0.19, 0.22}
\definecolor{radicalred}{rgb}{1.0, 0.21, 0.37}

\hypersetup{ linktoc=all,
    colorlinks, linkcolor={palatinateblue},
    citecolor={brightpink}, urlcolor={amaranth}
}
\newcommand{\changeurlcolor}[1]{\hypersetup{urlcolor=#1}}
\graphicspath{{Images/}}

\makeatletter
\def\@fnsymbol#1{{\ifcase#1\or \textleaf  \else\@ctrerr\fi}}
	\makeatother

\begin{document}

\title{Zero Mass Remnant as an Asymptotic State of Hawking Evaporation}

\author{Yen Chin \surname{Ong}}
\email{ycong@yzu.edu.cn}
\affiliation{Center for Gravitation and Cosmology, College of Physical Science and Technology, \\Yangzhou University, Yangzhou 225009, China}
\affiliation{School of Physics and Astronomy,
Shanghai Jiao Tong University, Shanghai 200240, China}

\begin{abstract}
We show that although generalized uncertainty principle with negative parameter allows black holes to evaporate completely, this process takes an infinite amount time to achieve, resulting in a metastable remnant.
\end{abstract}

\maketitle

\section{Generalized Uncertainty Principle: Positive vs. Negative Correction}\label{Intro}

The generalized uncertainty principle (GUP) is a quantum gravity inspired correction to the Heisenberg's uncertainty principle. The simplest form of GUP is
\begin{equation}
\Delta x \Delta p \geqslant \frac{1}{2} \left[\hbar + \frac{\alpha L_p^2 \Delta p^2}{\hbar}\right],
\end{equation}
where $L_p$ denotes the Planck length. The GUP parameter $\alpha$ is typically considered to be a positive number of order unity in theoretical calculations.
Note that GUP is largely heuristically ``derived'' from Gedanken-experiments and is often taken as a phenomenological model \cite{9301067, 9305163, 9904025, 9904026}.
 
GUP with positive $\alpha$ leads to a black hole remnant  because the generalized Hawking temperature for an asymptotically flat Schwarzschild black hole takes the form \cite{0106080}
\begin{equation}\label{1}
T[\alpha>0] = \frac{Mc^2}{4\alpha \pi} \left(1-\sqrt{1-\frac{\alpha \hbar c}{GM^2}}\right),
\end{equation}
and temperature has to be a real number. Thus the black hole stops evaporating when it reaches mass $M = (\alpha \hbar c/G)^{1/2}=\sqrt{\alpha}M_p$, which is of the order of a Planck mass for $\alpha = O(1)$. Remnants \cite{remnant} are problematic (see review \cite{1412.8366}), but having a finite temperature object as an end state of Hawking evaporation is arguably more palatable than having a divergent temperature as in the usual picture of Hawking evaporation (assuming that $T \propto 1/M$ holds for all time\footnote{While this is commonly assumed, various approaches exist which modify the behavior at late time and thereby avoiding divergence in the temperature, e.g., by considering microcanonical ensembles \cite{9712017,  0004004, 0412265, 1101.1384}.}). This remnant picture is peculiar -- it has finite temperature (which is quite high for $M \sim M_p)$, but it is thermodynamically inert (as can be shown, its specific heat is zero, see also Sec.(\ref{Discussion})) -- i.e., it no longer interacts with the environment thermally. Instead, it behaves like an elementary particle, and so the temperature here is better interpreted as the energy of the particle ($E=k_BT$). 

Nevertheless, GUP with positive $\alpha$ allows white dwarfs to be arbitrarily large \cite{1512.06356,1512.04337}, which is observationally problematic.
If one takes $\alpha < 0$ however, the Chandrasekhar limit is restored, while at the same time there is no divergence in Hawking temperature at late time \cite{1804.05176}. There is previously very little discussion on the possibility of negative $\alpha$, but see \cite{1407.0113} by Scardigli and Casadio, in which $\alpha<0$ is derived by assuming that the GUP-corrected Hawking temperature can be obtained from Wick-rotating the effective Schwarzschild-like metric with 
\begin{equation}
g_{tt}=-\left(1-\frac{2M}{r}+\varepsilon \frac{M^2}{r^2}\right);
\end{equation} 
obtaining
$\alpha = -4\pi^2\varepsilon^2\left[{M}/(2 M_p)\right]^2 < 0.$

\begin{figure}[!h]
\centering
\includegraphics[width=3.3in]{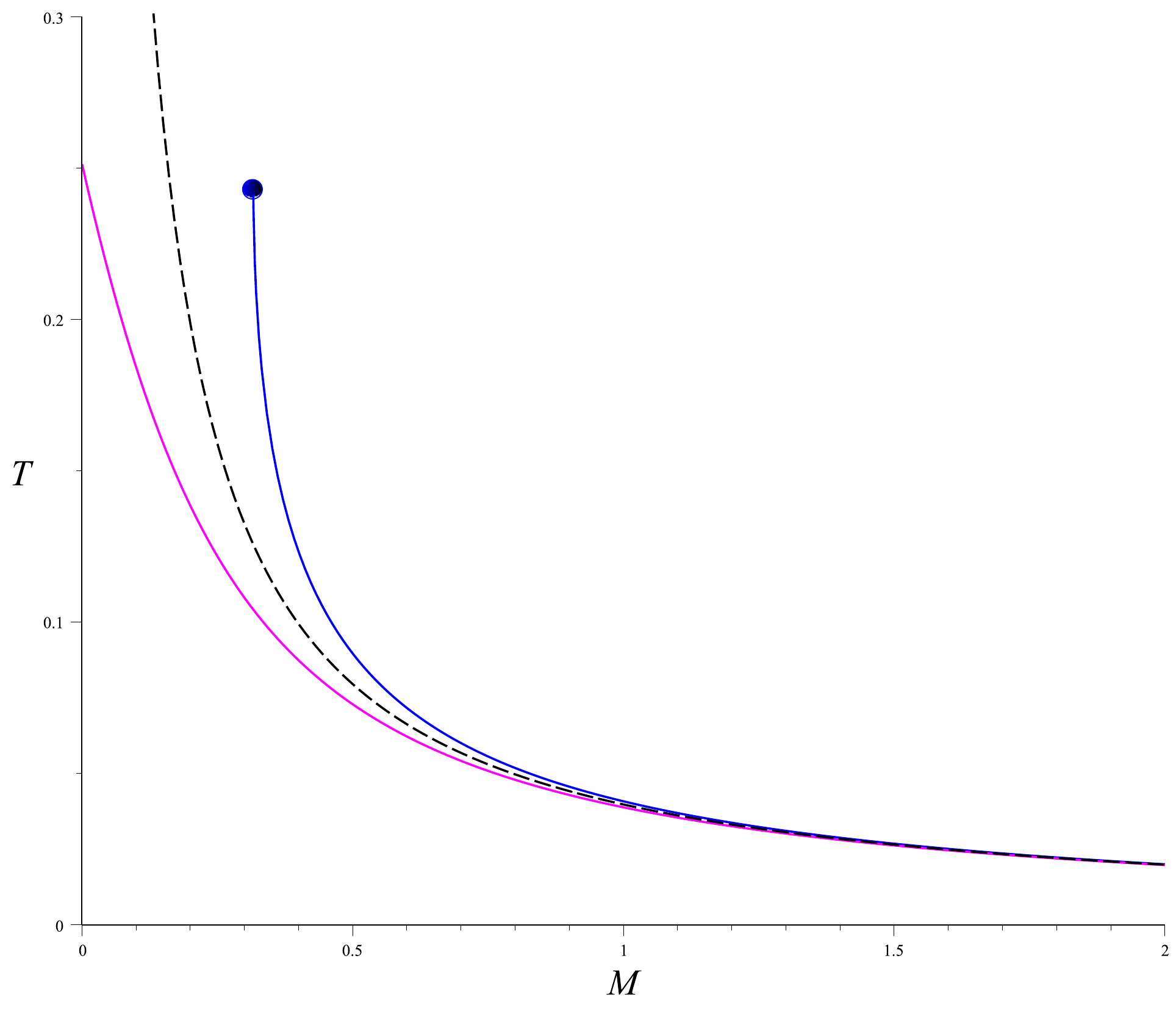}
\caption{The various Hawking temperature of an asymptotically flat Schwarzschild black hole. The usual Hawking temperature corresponds to the the middle dashed curve, which diverges as $M \to 0$. The infinity is ``cured'' with GUP correction: if $\alpha > 0$,  the temperature curve terminates at around $M \sim \sqrt{\alpha}M_p$, as shown by the right-most curve. If $\alpha < 0$, GUP correction no longer imposes a lower bound on the black hole size. This corresponds to the left-most curve: the temperature remains finite as the black hole shrinks down to zero size.
\label{fig6}}
\end{figure}
 
Explicitly, we can re-write Eq.(\ref{1}) as
\begin{equation}
T[\alpha<0] = -\frac{Mc^2}{4|\alpha| \pi} \left(1-\sqrt{1+\frac{|\alpha| \hbar c}{GM^2}}\right).
\end{equation}
There is now, strictly speaking, no remnant, since the mass of the black hole can go to zero when temperature reaches $T  = 1/\sqrt{4\pi|\alpha|}$. See Fig.(\ref{fig6}). In other words, the black hole dissolves into particles of said temperature. We shall refer to the $\alpha > 0$ and $\alpha < 0$ cases as, respectively, ``positive correction'' and ``negative correction'' to the uncertainty principle. 

In \cite{0912.2253}, Jizba-Kleinert-Scardigli referred to such the end state of $\alpha < 0$ GUP corrected black hole as a ``remnant with zero rest mass''. As we shall see below, this turns out to be an apt description.

In this work, we further investigate the evolution of negatively corrected GUP black holes. We show that the black hole takes infinite amount of time to evaporate down to $M=0$, and so for all practical purposes it is a metastable, long-lived, remnant. From here onwards, we will set $G=c=\hbar=k_B=1$. 

\section{The Lifetime of GUP-Corrected Black Holes}

\begin{figure}[!h]
\centering
\includegraphics[width=3.4in]{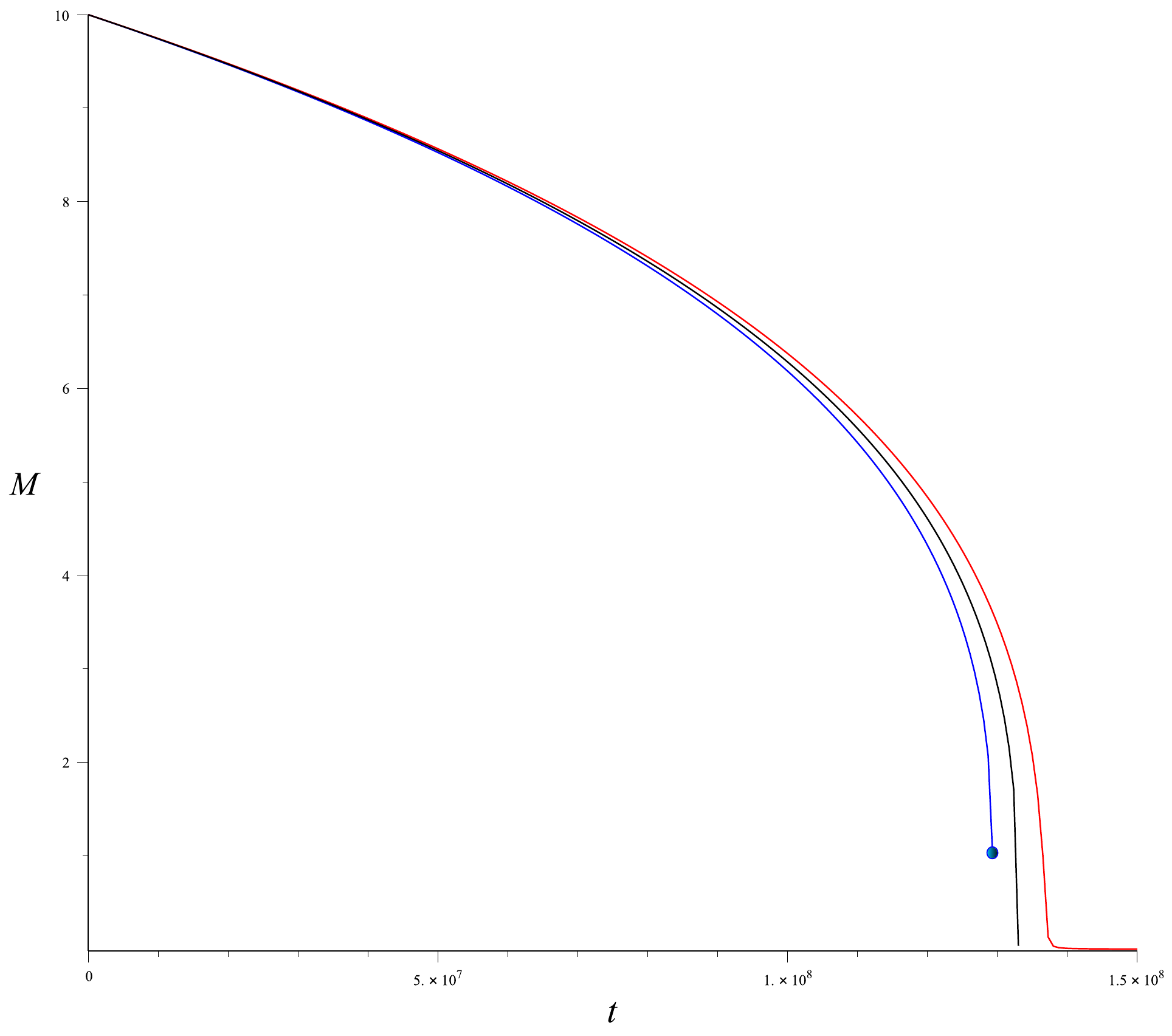}
\caption{Mass evolution of Schwarzschild black holes with no GUP correction (black, middle curve), positive GUP correction (blue, left curve), and negative GUP correction (red, right curve).
The positive GUP correction leads to a remnant in finite time, while negative GUP correction yields infinite lifetime, as the black hole slowly evaporates indefinitely. These contrast with the usual case without GUP correction, in which the black hole completely evaporates in finite time.
\label{M-t}}
\end{figure}

\begin{figure}[!h]
\centering
\includegraphics[width=3.4in]{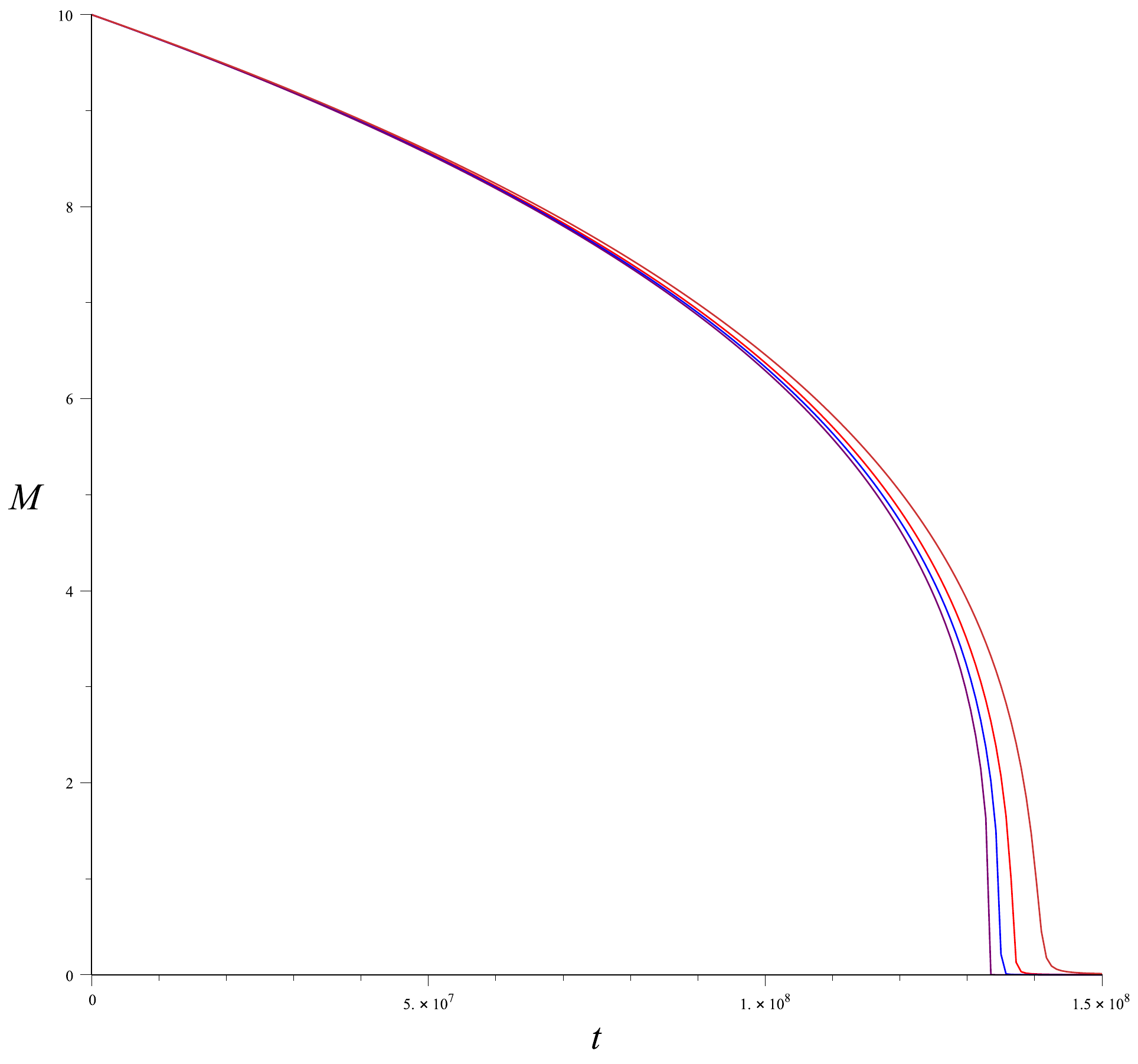}
\caption{Mass evolution of Schwarzschild black holes with negative GUP corrections. From right to left, the curves correspond to GUP parameters $\alpha=-2,-1,-0.5,-0.1$, respectively. Decreasing $\alpha$ (i.e. more negative) has the effect of pushing the curve towards the right. They all asymptote towards zero.
\label{M-t-2}}
\end{figure}

In the usual picture of an asymptotically flat Schwarzschild black hole evaporation, we have the thermal evolution
\begin{equation}\label{ODE}
\frac{\text{d}M}{\text{d}t} = -\alpha \sigma aT^4,
\end{equation}
where $a$ is the radiation constant ($4/c$ times the Stefan–Boltzmann constant), $\sigma$ is essentially the area (in the geometric optics approximation, it is the photon capture cross section $\sigma = 27\pi M^2$), and $\alpha$ the greybody factor. For our purpose we are only interested in the general and qualitative behavior of the evolution, so we will simply consider the ODE
\begin{equation}\label{thermal}
\frac{\text{d}M}{\text{d}t} = -\frac{1}{(8\pi)^4M^2},
\end{equation}
since $\sigma \sim M^2$, and $T=1/(8\pi M) = T[\alpha=0]$. We retain the factor $8\pi$ in the Hawking expression so as to be consistent with the GUP temperature expression (in the limit $\alpha \to 0$).
It is well-known that Schwarzschild black hole completely evaporates in a finite time. In Fig.(\ref{M-t}), we show the numerical plot (setting initial mass $M_0=10$). The black hole evaporates at about $t_\text{evap}= 1.293 \times 10^9$. If we replace the temperature expression with Eq.(\ref{1}) instead, 
we have the GUP-modified ODE:
\begin{equation}\label{ODE2}
\frac{\text{d}M}{\text{d}t} = -M^2 \cdot \left(\frac{M}{4\alpha \pi}\right)^4 \left(1-\sqrt{1-\frac{\alpha }{M^2}}\right)^4.
\end{equation}
We find that a positively corrected GUP black hole with $\alpha=1$ stops evaporating at a time somewhat less than $t_\text{evap}$, but of the same order of magnitude. Our main interest is to investigate what happens when $\alpha < 0$. Take, for example, $\alpha=-1$. The result is intriguing: the black hole evaporation is divided into two regimes: in the first stage, it follows the same qualitative behavior as that of $\alpha=0$ and $\alpha>0$ black holes. However, just as it approaches the horizontal axis, it turns around and asymptotes along the $t$-axis. In fact, it never truly reaches zero mass. See Fig.(\ref{M-t}). Decreasing $\alpha$ (i.e. more negative) has the effect of pushing the curve towards the right, as shown in Fig.(\ref{M-t-2}).

To appreciate the infinite lifetime of such black holes, we note that for negative $\alpha$, the mass evolution follows
\begin{equation}
\frac{\text{d}M}{\text{d}t} = \frac{M^6}{(4|\alpha| \pi)^4} \left(1-\sqrt{1+\frac{|\alpha| }{M^2}}\right)^4.
\end{equation}
When $M$ has become sufficiently small, we have
\begin{equation}
\frac{\text{d}M}{\text{d}t} \sim -\frac{M^2}{(4\pi)^4 {\alpha^2}}.
\end{equation}
This has solution of the form
\begin{equation}
M = M_0\left(\frac{256\pi^4 \alpha^2}{256\pi^4 \alpha^2 + M_0t}\right),
\end{equation}
where $M_0$ is the ``initial'' (small) mass. Clearly $M \to 0$ as $t \to \infty$.

\section{Sparsity Makes Lifetime Even Longer}

Hawking temperature is unlike a typical blackbody: the wavelength of a Hawking particle (without GUP correction) is
\begin{equation}
\lambda = \frac{2\pi}{T} = 16\pi^2 M \approx 79 r_h \gg r_h,
\end{equation}
where $r_h = 2M$ is the horizon. This is in contrast with a typical blackbody radiation, e.g., from a burning coal, which has wavelength much smaller than the size of the emitting body.
For black holes, the emitting surface is taken to be the geometric optics cross section $\sigma$. One can look at the dimensionless ratio\footnote{We have omitted a numerical prefactor in the definition of $\eta$, which plays no role in our discussion below.}
$
\eta := {\lambda^2}/{\sigma}.
$

If $\eta \ll 1$, then we have a typical blackbody, which emits continuously. On the other hand, $\eta \gg 1$ means the Hawking radiation is extremely \emph{sparse}: a particle is randomly emitted in a discrete manner, with ``pauses'' in between. (Such emission also contributes to the random walk of black hole, as the result of backreaction from Hawking emission \cite{random}.) See \cite{1506.03975, 1512.05809, 1606.01790} for detailed discussions. 

A recent study investigated the effect of GUP with positive $\alpha$ on the sparsity of Hawking radiation, and found that the radiation is no longer quite sparse near the Planck scale \cite{1801.09660}. We will re-do the calculations as there are some minor disagreements with \cite{1801.09660}, which nevertheless do not affect their results. We then consider the case of negative $\alpha$, and found that it enhances the sparsity -- in fact, the radiation becomes \emph{infinitely sparse}. Note that the usual calculation of the mass loss rate does not take into account the effect of sparsity -- mass loss is simply treated as particle mass loss associated to a given temperature \cite{1506.03975, 1512.05809, 1606.01790}. However, once the wave effect of the radiation is taken into account, sparsity becomes an important feature that extends the black hole lifetime. For our negatively corrected GUP black hole, its lifetime is already infinite \emph{before} taking into account sparsity. In addition, we have not taken into account greybody factors, which would suppress the rate even further. Therefore it is safe to conclude that these black holes do have infinite lifetime (unless new physics comes into effect). Now let us look at the details of the calculations.

\begin{figure}[!h]
\centering
\includegraphics[width=3.4in]{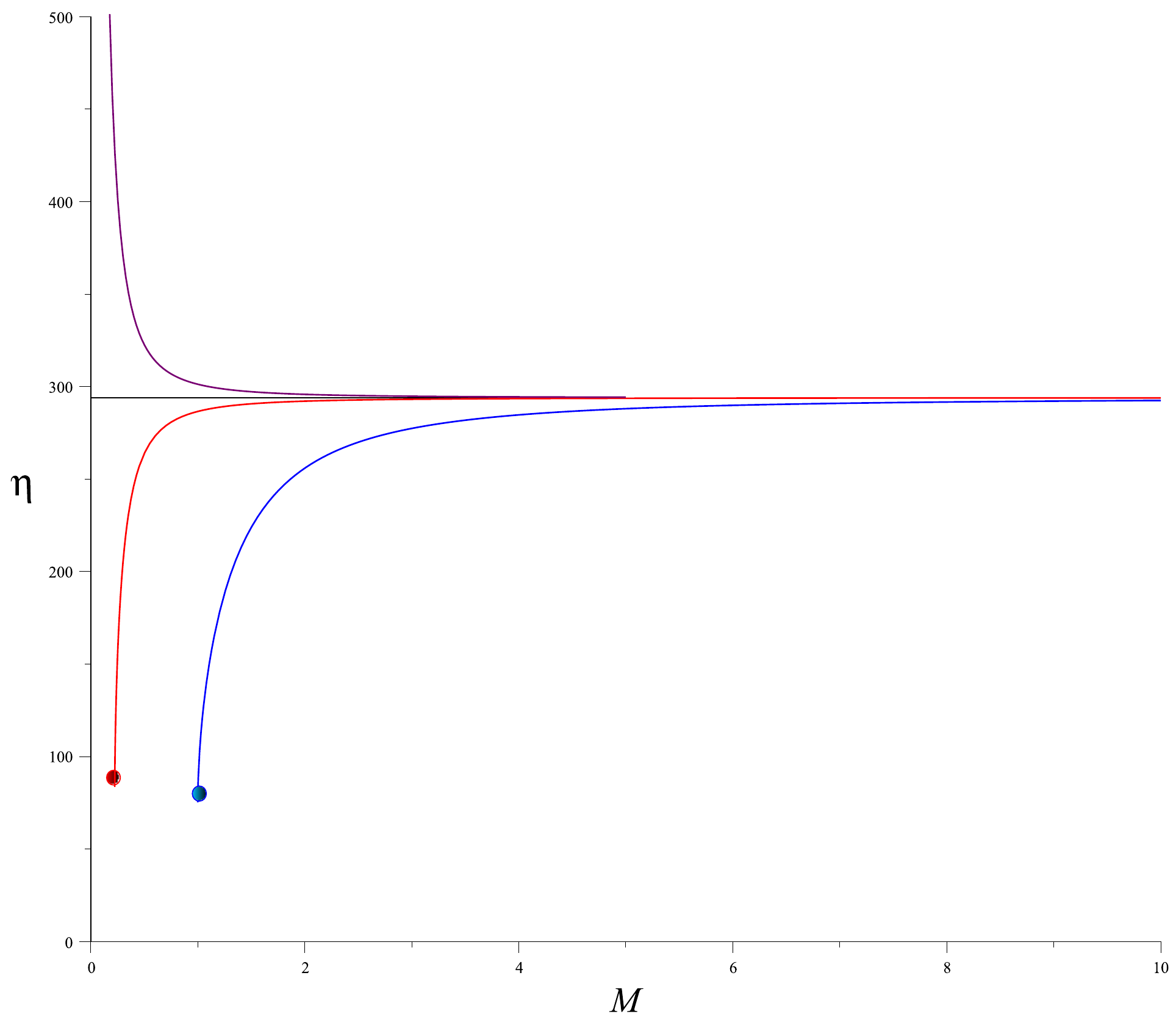}
\caption{The sparsity of Hawking radiation. The constant black line corresponds to the usual picture of Hawking evaporation -- the radiation remains sparse even when the temperature is high towards the end of the evaporation. The bottom two curves are for $\alpha=1$ (bottom right curve, blue) and $\alpha=0.05$ (bottom left, red), respectively. Indeed, for $\alpha > 0$: GUP correction leads to the decrease in $\eta$, and therefore the radiation becomes less sparse towards the end. However, for $\alpha < 0$, we get an ever-increasing $\eta$, which implies that the radiation becomes \emph{extremely} sparse. Shown here (top, purple), is an example for which $\alpha=-0.05$.
\label{s}}
\end{figure}

After GUP correction (with $\alpha$ of either signs), we have 
\begin{equation}
\lambda [\alpha] = \frac{2\pi}{T[\alpha]}.
\end{equation}
Thus, GUP corrected sparsity is
\begin{equation}\label{GUPs}
\eta [\alpha] := \frac{\lambda^2[\alpha]}{\sigma}=\frac{\lambda^2[\alpha]}{27\pi M^2}.
\end{equation}
In \cite{1801.09660}, the authors considered the fact that $\sigma=(27/4)\pi r_h^2$. Denoting $A=\pi r_h^2$, they then considered a GUP-correction to $A$ essentially by treating $S= A/4$ following Bekenstein-Hawking entropy, and the fact that $S$ is modified under GUP correction via the first law of black hole thermodynamics:
\begin{flalign}
S[\alpha]&=\int{\frac{1}{T[\alpha]}}\text{d}M \notag \\&= 2\pi \left[M^2 + M\sqrt{M^2-\alpha} - \alpha \ln(M+\sqrt{M^2-\alpha})\right] \notag \\&~~~+ \text{const.}
\end{flalign}
Note that there is a logarithmic term correction to the entropy. However, in the derivation of the GUP-Hawking temperature itself, it is usually considered that the horizon is still at $r=2M \sim \Delta x$, so that to be self-consistent, the horizon size should not receive any GUP correction\footnote{More explicitly, in the heuristic derivation of Hawking temperature via the usual Heisenberg's uncertainty principle, one takes $\Delta x \sim 2M$, and $T \sim \Delta p \sim 1/\Delta x$. In the GUP case, one replaces $\Delta p$ with the GUP corrected version, but  $\Delta x$ remains $2M$ \cite{0106080}.}. As a consequence, the area law $S=A/4$ does \emph{not} hold\footnote{This is consistent with other approaches of quantum gravity in which logarithmic correction appears in the entropy expression, but one usually does not modify the area expression. See, e.g., \cite{0401070,1112.3078}.}. Hence, we choose to not consider correction to the horizon size (for the same reason, the $\sigma$ in Eq.(\ref{ODE2}) remains just $\sigma \sim M^2$, this agrees with \cite{0205106, 0406514}). Despite this difference, the result for $\alpha >0$ remains qualitatively the same as \cite{1801.09660}, namely $\eta$ decreases as the black hole mass gets near the Planckian regime\footnote{We note that a previous study without utilizing GUP, but taking into account backreaction instead, leads to the same conclusion \cite{1601.07310}, including the existence of a remnant.}, see Fig.(\ref{s}). Another small difference with \cite{1801.09660} is a technical one: they have used series expansion in the temperature, while we retained the full expression. This is the reason why their black holes mass can become zero, while we know that for $\alpha > 0$ there should be a minimum mass corresponds to the remnant mass.
Interestingly, as shown in Fig.(\ref{s}), for $\alpha < 0$, the sparsity increases without bound. It is easy to see that it diverges: $M$ can tend to zero for $\alpha < 0$, while $T$ tends to a constant value $1/\sqrt{4\pi |\alpha|}$, so from Eq.(\ref{GUPs}), $\eta \to \infty$.

\section{Discussion: The Final Fate of GUP-Corrected Black Holes}\label{Discussion}

What happens to an evaporating black hole as its mass gets sufficiently small has always been a puzzle in theoretical physics. A common belief is that black holes should completely evaporate away, since there are many problems with the remnant proposal, as discussed in the review paper \cite{1412.8366}. Of course, the latter option, though unpopular, still deserves a closer scrutiny.  

Indeed, there are many reasons to suspect that the simple thermal description of black hole, i.e. the ODE in Eq.(\ref{ODE}),
cannot hold for all scales. In addition to micro-canonical ensemble considerations \cite{9712017, 0004004, 0412265, 1101.1384}, from quantum information point of view, it is widely believed that unitarity requires information to ``leak'' out of the black hole after Page time \cite{page1, page2} (though in highly scrambled form among the quantum entanglement of the Hawking particles) -- about half-way through the evaporation. Subtle quantum effects might affect the evolution of black holes. Furthermore, even without these effects, as the black hole becomes hotter and hotter, it is likely that new physics would come into effect as one reaches sufficiently high energy scale. The evolution of black holes might then get modified. In other words, perhaps quantum gravity will resolve the divergence in the temperature, much like how it might resolve black hole singularities. 

GUP is a quantum gravity inspired phenomenological model that could accommodate minimal length (for positive GUP parameter) and black hole remnant, see review \cite{1203.6191}, so it is interesting to investigate how black holes evolve under GUP-corrected Hawking process. In this work, we explore the possibility that the GUP parameter $\alpha$ is negative, a possibility that  is much less studied. We found that despite the absence of lower bound for the mass, complete evaporation cannot be achieved in finite time. In addition, sparsity of the Hawking radiation becomes infinite in the $M \to 0$ limit. Considerations of greybody factors will likely  ``prolong'' the (already infinite) lifetime.  Therefore the end state of the black hole should be viewed as a ``zero mass remnant'' -- in the sense that it is a metastable state that asymptotes to zero mass. 

Lastly we remark on the heat capacity of GUP-corrected asymptotically flat Schwarzschild black holes. Heat capacity is defined by
\begin{equation}
C:=\frac{\text{d}M}{\text{d}T}=\frac{\text{d}M}{\text{d}t}\left(\frac{\text{d}T}{\text{d}t}\right)^{-1}.
\end{equation}
It can be shown that $C[\alpha]\leqslant 0$ for all values of $\alpha$, just like the usual Schwarzschild black hole. However, for $\alpha > 0$ case, as the remnant mass is approached, $C \to 0$. Thus, a remnant has zero heat capacity -- it is thermodynamically inert, as explained in Sec.(\ref{Intro}). For $\alpha < 0$ however, heat capacity is always negative, so the metastable remnant interacts thermally with the environment, as already pointed out in \cite{0912.2253}.

It is interesting how different signs of $\alpha$ lead to black hole remnants: for $\alpha > 0$, it is a stable remnant of finite ``temperature'' (best interpreted as the energy of the remnant ``particle'', see Sec.(\ref{Intro})); whereas for $\alpha < 0$, it is a metastable, long-lived remnant that approaches zero rest mass asymptotically.  
The latter cannot be directly inferred from the $T-M$ plot alone (Fig.(\ref{fig6})), one must instead study its evolution equation and sparsity.

The implications for such a remnant deserve a closer study. We speculate on some possibilities: firstly, remnants could play some roles in ameliorating the information paradox (see \cite{1412.8366}). Furthermore, it was proposed in \cite{1506.03975} that the sparsity in Hawking radiation could encode some information between the emission gap without disturbing the time-averaged emission spectrum. Although this is unlikely to be the full resolution to the information paradox, it would be interesting to investigate in details the quantum information and entanglement properties of the Hawking particles emitted by our long-lived remnant with infinite sparsity towards the end of time. Secondly, black hole remnants, including those obtained from positively corrected GUP \cite{0205106, 0406514} , had been considered as candidate of dark matter -- see, e.g., the references in \cite{1805.03872}. Remnants as dark matter is still allowed by observational constraints \cite{1607.06077}.
One could look into how phenomenologically different the zero mass remnant might be as dark matter candidate, compared to the $\alpha > 0$ remnant.

\begin{acknowledgments}
YCO thanks the National Natural Science Foundation of China (grant No.11705162) and the Natural Science Foundation of Jiangsu Province (No.BK20170479) for funding support. 
He also acknowledges the China Postdoctoral Science Foundation (grant No.17Z102060070) for partial support.
YCO thanks Brett McInnes for comments and suggestions. 
\end{acknowledgments}

\end{document}